\documentclass[final,authoryear,5p,times,twocolumn]{elsarticle}

\usepackage{graphics}
\usepackage{amssymb}
\usepackage{epigraph}

\usepackage{lineno}

\journal{Planetary and Space Science}

\begin{document}

\twocolumn[
\section*{\Large {\mdseries Editorial}}
\section*{\LARGE {\sc Cosmic Dust \MakeUppercase{\romannumeral 9}}}
\begin{flushleft}
\hrulefill
\end{flushleft}
\vspace{0.3in}
]

{\sl \epigraph{For dust you are and to dust you will return}{Genesis 3:19}}

\section{Interdisciplinary research on cosmic dust}

%\section{Introduction}
%\label{introduction}

%{\sl \epigraph{For dust you are and to dust you will return}{Genesis 3:19}}

It is a dusty universe. 
Every year, approximately $10^4~\mathrm{tons}$ of cosmic dust enter Earth's atmosphere \citep{brownlee1985}. 
These dust particles mostly originate from the breakup of comets and asteroids and sometimes are visible as meteors if they are greater than millimeters. 
Besides originating from comets and asteroids, interplanetary dust reveals itself through the zodiacal light, caused by sunlight scattered at the particles. 
The collisional breakup of asteroidal bodies and outgassing of cometary bodies are also considered as the major sources of dust particles in debris disks around main sequence (MS) stars, revealed through the detection of infrared (IR) emission and/or scattered optical/near-IR light. 
Known as protoplanetary disks, dusty disks are also seen in pre-MS Herbig Ae/Be stars and T Tauri stars. 
Moreover, dust is associated with stars in their late stages of evolution: dust grains condense in the atmospheres and blown out of cool evolved stars by radiation pressure; dust grains also condense in the cool ejecta of supernovae. Dust pervades the interstellar medium (ISM) of the Milky Way galaxy and external galaxies, both near and far. 
There is evidence for the presence of an obscuring dust torus around the central engine of active galactic nuclei (AGNs). Dust grains are also present in the intergalactic space. 

\begin{figure*}[t]
 \begin{center}
  \includegraphics[width=180mm]{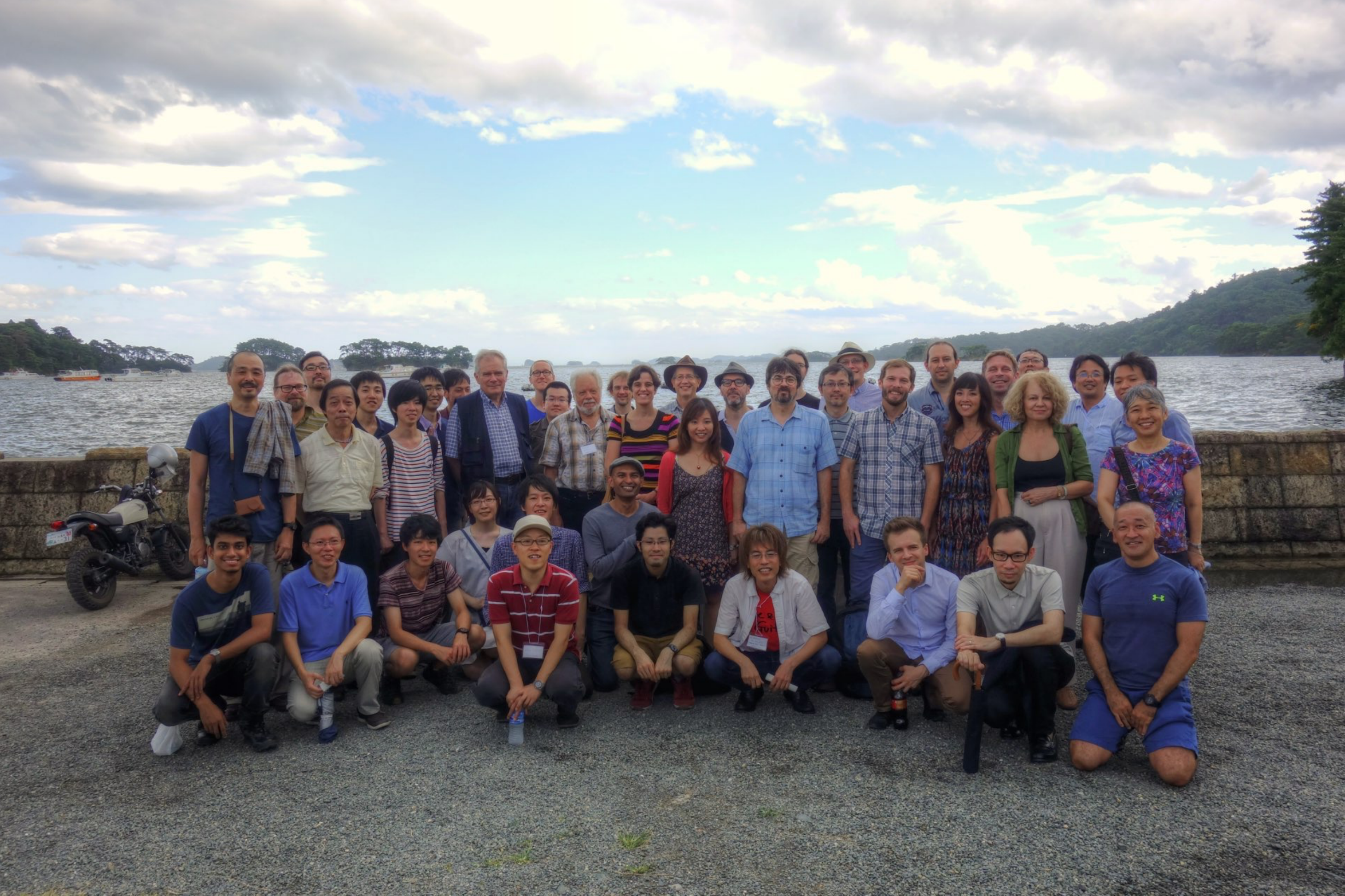}
 \end{center}
 \caption{A group picture of participants to {\sc Cosmic Dust \MakeUppercase{\romannumeral 9}}; (In no particularly order) 
T. Fujiyoshi, C.~M. Wright, H. Kobayashi, T. Onaka, T. Kokusho, M. Nashimoto, S. Ishiki, T. Kimura, J. Kre\l owski, K. Penner, K. Sano, N.~N. Kiselev, F. Le Petit, C. Kemper, J.~M. Bauer, R. Wu, P. Thebault, G.~C. Sloan, P. Scicluna, T. Ootsubo, J.~E. Lindberg,  F. Galliano, A. Nguyen, A. Boogert, T. Nozawa, L. Kolokolova, H. Senshu, T. Shimonishi, A.~M. Nakamura, G. Saikia, X. Zhang, T. Nagao, T. Omura, A.~K. Inoue, T. Ueda, Z. Wahhaj, R. Tazaki, H. Kimura, B. Safonov, T. Hirai, and H. Chihara.}
 \label{fig:one}
\end{figure*}

It has long been recognized that dust is a ubiquitous feature of the cosmos, impinging directly or indirectly on most fields of modern astronomy. 
Aiming at providing a constructive forum for representatives of a variety of diverse fields in dust astrophysics to lively discuss their latest results in a cozy atmosphere, we have been organizing a series of cosmic dust meetings since 2006.  
The 9th meeting on this series (hereafter {\sc Cosmic Dust \MakeUppercase{\romannumeral 9}}) was held at the Aoba Science Hall of Tohoku University in Sendai, Japan on August 15--19, 2016. 
It was attended by 51 participants from 12 countries (see Fig.~\ref{fig:one} for a group picture taken at Matsushima, one of the Three Views of Japan). 

The invited speakers, James M. Bauer (JPL, USA), Adwin Boogert (Sofia Science Center, USA), Hiroki Chihara (Osaka Sangyo University, Japan), Fr\'{e}d\'{e}ric Galliano (CEA/Saclay, France), Akiko M. Nakamura (Kobe University, Japan), Ann Nguyen (JETS/Jacobs Tech., USA), Ralf Siebenmorgen (ESO, Germany), Greg C. Sloan (Cornell University, USA), Philippe Thebault (LESIA, France), and Christopher M. Wright (UNSW ADFA, Australia), each presented a 40-minute talk. 
In addition, there were also 30 contributed talks of which each was assigned 20 minutes and 13 posters all of which were displayed throughout the entire 5-day meeting. 
The invited speakers ranked the posters and selected Takuma Kokusho (Nagoya University, Japan) and Ryo Tazaki (Kyoto University, Japan) as the laureates of the Best Poster Award for their posters ``A systematic study on dust in early-type galaxies'' and ``Grain alignment in the protoplanetary disks'', respectively. 
The invited talks, contributed talks, and posters covered a wide variety of topics, ranging from dust in the Solar System to dust in young stars, MS stars, and evolved stars as well as dust in the Galactic and extragalactic ISM and in AGNs.

This special issue in Planetary and Space Science (PSS) is a collection of (some of) the papers resulted from the meeting except for \citet{pitman-et-al2017} and \citet{rachid-et-al2017}. 
It consists of 5 review articles and 7 original papers. 
Each paper was peer reviewed by two or more anonymous referees. 
We thank all the authors, referees and editors whose tireless efforts made it possible for this special issue to be a valuable source of dust research.  

\section{The contents of {\sc \bf Cosmic Dust \MakeUppercase{\romannumeral 9}}}
\label{review}

\subsection{Debris disks and Solar System}

Studies on the physics in collision between dust particles hold the key to understanding the evolution of cold debris disks currently observed around MS stars.
As pointed out by Philippe Thebault, modeling of debris disks relies largely on the knowledge of collisional physics that is valid for meter-sized or larger objects \citep{thebault2016}.
Therefore, it is of great importance that the results of laboratory experiments on collisional phenomena are well explained on the basis of theoretical arguments. 
Akiko M. Nakamura provided a comprehensive review on impact processes and outcomes of porous materials in the strength-dominated regime from experimental and theoretical points of view \citep{nakamura2017}.
It is, however, unclear whether extrapolation of the collisional physics to micrometer-sized grains is justified, since a lack of cracks in micrometer-sized and smaller grains could enhance the strength of the grains \citep{chokshi1993}.
{\sl As a consequence, we caution that the production and destruction of small dust grains in debris disks and the Solar System might have been overestimated unless microphysics is considered in the modeling.}

%Other papers covered the following topics: ... by Hiroshi Kobayashi; ... by Zahed Wahhaj; ... by Hiroshi Kimura.

%\subsection{Solar System}

The origin of dust in the zodiacal cloud is still a matter of debate, but asteroids and short-period comets are the major sources to release dust particles near the ecliptic plane.
James (Gerbs) Bauer reviewed the results obtained with the NEOWISE mission, in which he is the Deputy Principal Investigator \citep{bauer2016}.
The mission provided a full-sky survey of Near Earth Objects, among them there were 164 comets at two wavelengths of $\lambda = 3.4$ and $4.6~\mathrm{\mu m}$.
The NEOWISE observations of dust particles in cometary trails have shown that for all types of comets, the size of the particles appeared to lie mainly in a millimeter-size range, and most of them were emitted from their parent bodies near perihelion.
The results were well explained by the dynamics of dust particles released from their parent bodies, because dust particles smaller than millimeters are expelled by solar radiation pressure and dust particles larger that millimeters are too heavy to be emitted without intense activities near perihelion.
Takahiro Ueda reported that such millimeter-sized particles could produce the asymmetry of the zodiacal light observed by AKARI in the mid-infrared, while smaller particles are the main contributors to the zodiacal light \citep{ueda-et-al2017}.
Coronagraphic images of debris disks have also revealed the presence of asymmetries in the disks at projected distances in which Kuiper-belt objects could reside \citep{kalas-jewitt1995}.
{\sl A detailed study on dust particles in the zodiacal light could provide a clue to the source and sink of dust particles in debris disks observed around main-sequence stars.}

Other papers covered the following topics: New insights into gas-induced trapping of small silicate grains in debris disks by Hiroshi Kobayashi; the first H-band resolved image of a narrow dust ring around a 16~Myr-old star by Zahed Wahhaj; the importance of grain surface curvature on photoelectric yields by Hiroshi Kimura; an experimental study on the mechanical strengths of chondrules and chondrule-analogs by Sae Shigaki (presented by A.~M. Nakamura); a new interpretation of the electric currents measured by Lunar Dust Experiment (LDEX) onboard LADEE by Xiaoping Zhang; the effect of high-energy photoelectrons on the dynamics of charged grains by Hiroki Senshu; mutual contaminations of gas and dust in the measured polarization of cometary comae by Nikolai N. Kiselev; a summary of Rosetta's in-situ dust measurements by Ludmilla Kolokolova; far-infrared observations of zodiacal dust bands with the AKARI satellite by Takafumki Ootsubo; an experimental study on the variation of powder porosity with compression by Tomomi Omura \citep{omura-nakamura2017}.

\subsection{Evolved stars, supernovae, and star-forming regions}

Silicates are one of the abundant condensates formed in the atmospheres of red giants, AGB stars, supernovae, and novae, and then ejected into the ISM.
Ann Nguyen overviewed advanced nanoscale laboratory analysis of presolar silicate grains with diverse stellar origins and processing in the ISM and the Solar System \citep{nguyen2017}. 
It turned out that AGB stars dominate dust formation in space and the crystallinity of silicate grains in evolved stars is higher than that in the ISM.
This indicates that silicate grains are amorphized in the ISM, while the uniformity of interstellar silicate composition found by Cassini in-situ measurements could be associated with amorphization of silicate grains by evaporation and re-condensation in the ISM \citep[cf.][]{altobelli-et-al2016}.
{\sl Therefore, we cannot reject the possibility that isotropic anomalies of condensates in the atmospheres of evolved stars could be erased during homogenization of the grains in the ISM.}

Other papers covered the following topics: an overview of carbon-rich dust observed in evolved stars by Greg C. Sloan \citep{sloan2017}; a comprehensive review on the formation and thermal processing of ices in the Universe by Adwin Boogert; new insights into the origin of presolar grains based on numerical simulations of Galactic chemical evolution by Kenji Bekki; a numerical study on the effect of multiple scattering by circumstellar dust on extinction of light from bright point sources by Takashi Nagao; model constraints on supernovae-produced dust grains from ultraviolet extinction data by Mingxu Sun; an analysis of ice absorption features in the near- to mid-infrared spectra of AKARI toward candidates for young stellar objects (YSOs) by Tomoyuki Kimura; an improvement in the Meudon PDR (photodissociation region) code with the implementation of grain surface chemistry by Franck Le Petit; a study on gas-phase and grain-surface chemistry with a new time-dependent code by Johan E. Lindberg; an observational study on the physical conditions of dust and molecular gas at the PDR in the Great Nebula of Carina by Ronin Wu; a theoretical study on the physical conditions of supernova ejecta from the size of presolar corundum grains by Takaya Nozawa; data analysis of dust disks around pulsars in the wavelength range from near-infrared to submillimeter by Qi Li; hydrodynamic simulations on the motion of gas and dust under stellar radiation pressure inside H~II regions by Shohei Ishiki.

\subsection{Galaxies and interstellar medium}

Nearby galaxies are important targets for understanding grain evolution through the lifecycle of matter in the ISM.
They provide a much wider range of the ISM physical conditions than our Galaxy, such as the intensity and hardness of the interstellar radiation field, metallicity, stellar population, large-scale shocks and hot plasma.
Their proximity enables us not only to spatially resolve structures of galaxies but also to study faint galaxies such as dwarf and elliptical galaxies. 
Fr\'{e}d\'{e}ric Galliano presented a review on the properties of dust in nearby galaxies as seen by Herschel with the latest improvements in the dust spectral energy distribution (SED) and radiative transfer modelling \citep{galliano2017}. 
Recent observations indicate that the grain emissivity is larger by a factor of 2 to 3 than what was thought before in the far-IR and submillimeter region, which implies that we do not  correctly understand the composition of large interstellar grains. 
Herschel observations also show non-linear evolution of the dust-to-gas mass ratio with metallicity, suggesting grain growth with metal enrichment in a galaxy. 
Another puzzling issue is the submillimeter excess above the SED model prediction, which is conspicuous in dwarf galaxies. 
The excess appears to be associated with the diffuse gas phases, but its origin is uncertain; neither very cold dust, very large grains, nor fast spinning dust can probably explain the excess. 
{\sl Future sensitive far-IR to submillimeter spectroscopy with high spatial resolution would provide clues to these problems.} 

Other papers covered the following topics: an IR-emission modeling of dust mineralogy in AGNs and a radiative transfer modeling of silicate crystallinity in external galaxies by Ciska Kemper \citep{srinivasan-et-al2017}; a study on the determination of dust mass from submillimeter-wavelength observations by Peter Scicluna; a theoretical investigation of dust processing in elliptical galaxies by Hiroyuki Hirashita \citep{hirashita-nozawa2017}; a study on the spatial distribution of dust in galaxies by Kyle Penner; ALMA observations of dust formation in the early Universe by Akio K. Inoue; a report on the radio wave detection of a hot molecular core in the Large Magellanic Cloud by Takashi Shimonishi; a modeling study on the infrared silicate features of AGNs by Zhenzhen Shao; a systematic study on the IR observations of dust in early-type galaxies by Takuma Kokusho; a study on the interaction between galaxies from IR observations by Ayato Ikeuchi: a thorough modeling of the interstellar extinction and polarization of the diffuse ISM by Ralf Siebenmorgen \citep{siebenmorgen-et-al2017}; a discussion on the ubiquity of forsterite in embedded YSOs by Christopher M. Wright; the development of a numerical code for stochastic heating of small grains by Masashi Nashimoto; a discussion on the methods of measuring distances to distant objects by Jacek Kre\l owski \citep{krelowski2017}; a search for correlations between anomalous microwave emission from interstellar dust and AKARI data by Aaron C. Bell; AKARI observations of deuterated polycyclic aromatic hydrocarbons (PAHs) in the ISM by Takashi Onaka; near-IR observations of the diffuse Galactic light by Kei Sano; a study on the ultraviolet radiation scattered at dust particles around the Orion Nebula by Gautam Saikia \citep{saikia-et-al2017}.

\subsection{Protoplanetary disks}

IR spectral observations of dust particles in protoplanetary disks indicate that crystalline silicates are magnesium-rich and iron-poor \citep[e.g.,][]{olofsson-et-al2009}.
Hiroki Chihara has presented his laboratory measurements of IR spectra originating from amorphous silicates annealed at different temperatures and durations \citep{chihara-koike2017}.
His results suggest that annealing of amorphous silicate with a chondritic composition proceeds with an increase in the atomic abundance ratio of magnesium to iron.
Accordingly, he concluded that magnesium-rich, iron-poor crystalline silicates result from fractionation by annealing of amorphous silicates with a chondritic composition where iron remains in amorphous silicates.
However, amorphous silicates in the ISM appear to be magnesium-rich and iron-poor, implying that fractionation of amorphous silicates does not apply \citep{westphal-et-al2014,kimura2015,altobelli-et-al2016}.
Moreover, magnesium-rich, iron-poor crystalline silicates in the Solar System are condensates of solar nebula gas rather than annealed amorphous silicates \citep{brownlee-et-al2005,keller-messenger2011,keller-messenger2013}.
{\sl Therefore, the formation of magnesium-rich, iron-poor crystalline silicates does not necessarily require fractionation by annealing of amorphous silicates with a chondritic composition.}

Other papers covered the following topics: a report on the detection of silicon carbide in the protoplanetary disk of the close binary system SVS13 by Takuya Fujiyoshi; photometric and polarimetric visible observations of circumstellar metter around RW Aur A by Boris Safonov; molecular dynamics simulations on head-on collision between two nanometer-sized Ar particles by Hidekazu Tanaka; a theoretical estimate of timescales for grain alignment in protoplanetary disks by Ryo Tazaki.

\section{Perspectives for the development of cosmic dust research}
\label{perspectives}

A number of papers presented at {\sc Cosmic Dust \MakeUppercase{\romannumeral 9}} were devoted to processing of silicates such as amorphization, annealing, fragmentation, and compression \citep[e.g.,][]{nguyen2017,nakamura2017}.
A detailed investigation of the processing helps us to better understand the evolution of dust in the Universe, owing to the fact that silicates are the major refractory constituent of dust in a variety of cosmic environments.
Nonetheless, we have to confess that, in many cases, we are still far from full understanding of which processing of silicates is to a large extent at work.
\citet{hubbard2017} claim that thermal processing of silicates by FUors\footnote{FUors denotes variable T-Tauri stars with extreme outbursts named after the prototype low-mass pre-main sequence star FU Orionis.} outbursts is a key to the formation of rocky planetesimals and the architecture of planetary systems. 
In the last century, crystalline silicates in comets and debris disks were commonly believed to result from annealing of prestellar interstellar silicates in protoplanetary disks.
Nowadays, we know that comets are agglomerates of solar nebular condensates rather than presolar interstellar grains, but the most puzzling issue is which processing destroy prestellar interstellar grains in protoplanetary disks prior to the formation of planetesimals.
We could suppose that FUors outbursts are responsible for evaporation of prestellar interstellar grains and subsequent condensation of evaporated gas in protoplanetary disks.
This is, however, no more than a casual idea, indicating that we have to await further progress in cosmic dust research to answer the question.

%%% which interstellar dust is composed of either amorphous silicate or graphite.

%\section{Concluding remarks}
%\label{remarks}
The development of cosmic dust research is ongoing and the Cosmic Dust meeting series will provide a platform for discussion of all issues on cosmic dust.
Finally, the reader is cordially invited to take part in the Cosmic Dust meeting series and to join us for the development of cosmic dust research\footnote{Contact:~dust-inquries@cps-jp.org}. 
The 10th meeting on Cosmic Dust has already taken place in National Astronomical Observatory of Japan (NAOJ), Mitaka, Japan on August 14--18, 2017 and a special issue in PSS for this meeting is in preparation.
The 11th meeting will be held at JAXA/ISAS, Sagamihara, Japan on August 13--17, 2018. 
More information on the Cosmic Dust meeting series is available at the Cosmic Dust website (https://www.cps-jp.org/\~{}dust).

\section*{Acknowledgements}
%\label{acknowledgements}
We are grateful to the support of JSPS (Japan Society for the Promotion of Science), National Astronomical Observatory of Japan (NAOJ), Frontier Research Institute for Interdisciplinary Sciences, and Intelligent Cosmos Scientific Foundation, as well as the grant of Research Assembly supported by the Research Coordination Committee of NAOJ, National Institutes of Natural Sciences. 
We owe a special debt of gratitude to the LOC members: Takashi Shimonishi (Chair), Hiroki Chihara, Takayuki Hirai, Akio Inoue, Hiroshi Kobayashi, Takaya Nozawa, Tomomi Omura, Hiroki Senshu, Ryo Tazaki, and Koji Wada.

\bibliographystyle{model2-names}
\bibliography{<your-bib-database>}

%% not want to use model2-names.bst.


\begin{flushright}
Hiroshi Kimura\\
{\sl Graduate School of Science, Kobe University, Nada-ku Rokkodaicho 1-1, Kobe 657-8501, Japan}\\
{\sl Planetary Exploration Research Center (PERC), Chiba Institute of Technology, Tsudanuma 2-17-1, Narashino-shi, Chiba 275-0016, Japan}

\vspace{0.2in}
Ludmilla Kolokolova\\
{\sl Planetary Data System Group, Department of Astronomy, University of Maryland, College Park, MD 20742, USA}

\vspace{0.2in}
Aigen Li\\
{\sl 314 Physics Building, Department of Physics and Astronomy, University of Missouri, Columbia, MO 65211, USA}

\vspace{0.2in}
Hidehiro Kaneda\\
{\sl Graduate School of Science, Nagoya University, Furo-cho, Chikusa-ku, Nagoya 464-8602, Japan}

\vspace{0.2in}
Jean-Charles Augereau\\
{\sl Universit\'e Grenoble Alpes, IPAG, F-38000 Grenoble, France}\\
{\sl CNRS, IPAG, F-38000 Grenoble, France}

\vspace{0.2in}
Cornelia J\"{a}ger\\
{\sl Max Planck Institute for Astronomy, Heidelberg, Laboratory Astrophysics and Clusterphysics Group at the Institute of Solid State Physics, Friedrich Schiller University Jena, Helmholtzweg 3, 07743 Jena, Germany}
\end{flushright}

%\end{linenumbers}
\end{document}